\documentclass[aps,pre,superscriptaddress,showpacs,12pt]{revtex4}

\usepackage{amsmath}
\usepackage{graphicx}

\begin{document}

\title{Modulational instability of partially coherent signals in electrical transmission lines}


\author{M. Marklund}
\affiliation{Centre for Nonlinear Physics, Department of Physics, 
  Ume{\aa} University, SE--901 87 Ume{\aa}, Sweden}
  
\author{P. K. Shukla}
\affiliation{Centre for Nonlinear Physics, Department of Physics, 
  Ume{\aa} University, SE--901 87 Ume{\aa}, Sweden}
\affiliation{Institut f\"ur Theoretische
Physik IV and Centre for Plasma Science and Astrophysics,
Fakult\"at f\"ur Physik und Astronomie, Ruhr--Universit\"at
Bochum, D-44780 Bochum, Germany}

\date{Recieved at Phys.\ Rev.\ E February 24, 2006}


\begin{abstract}
  We present an investigation of the modulational instability of 
  partially coherent signals in electrical transmission lines.
  Starting from the modified Ginzburg-Landau equations and the
  Wigner-Moyal representation, we derive a nonlinear 
  dispersion relation for the modulational instability. It is found that 
  the effect of signal broadbandness reduces the growth rate of the modulational instability.
\end{abstract}

\maketitle


About a decade ago, Marquie {\it et al.} \cite{r1} and Bilbault {\it et al.} \cite{r2}
investigated theoretically and experimentally the nonlinear propagation of signals
in electrical transmission lines. Specifically, they considered a nonlinear network
composed of a set of cells containing two linear inductances in series and one in parallel,
together with a nonlinear capacitance diode in the shunt branch. It has been shown that the 
system of equations governing the physics of this network can be reduced to a cubic nonlinear 
Schr\"odinger (CNS) equation or a pair of coupled nonlinear Schr\"odinger (CNLS) equation. 
Both the CNS and the CNLS equations admit modulational instability and the formation of envelope 
solitons, which have been observed experimentally \cite{r1,r2}.

Recently, Kengne and Liu \cite{r3} presented a model for wave propagation on a discrete electrical
transmission line based on the modified complex Ginzburg-Landau (MCGL) equation, derived 
in the small amplitude and long wavelength limit using the standard reductive perturbation
technique and complex expansion \cite{r4} on the governing nonlinear equations. The MCGL is 
also referred to as the derivative nonlinear Schr\"odinger equation or the cubic-quintic 
Ginzburg-Landau equation. Nonlinear soliton solutions of the MCGL equation have been 
presented in  Ref. \cite{r3}.


In this Brief Report, we consider the modulational instabilities of partially coherent electrical 
pulses that are governed by the MCGL equation \cite{r3}
\begin{equation}\label{eq:nlse}
  i\partial_t u - P\partial_x^2u - \gamma u - Q_1|u|^4u 
    - iQ_2|u|^2\partial_xu - iQ_3\partial_x(|u|^2u) = 0,
\end{equation}
where $P$, $Q_j$ ($j=1,2,3$), and $\gamma$ are real transmission line  
parameters. We note that Eq.\ (\ref{eq:nlse}) has the space independent harmonic 
solution $u = u_0\exp(-i\Omega_0t)$, where $\Omega_0 = \gamma + Q_1U_0^4$. 

Next, we let $u = [u_0 + u_1(t,x)]\exp(-i\Omega_0 t)$ in Eq. (1), where $|u_1| \ll u_0$, and 
linearize with respect to $u_1$ to obtain 
\begin{eqnarray}
   i\partial_t u_1 - P\partial_x^2u_1 - 2Q_1u_0^4(u_1 + u_1^*)
    - iQ_2u_0^2\partial_xu_1 - iQ_3u_0^2\partial_x(2u_1 + u_1^*) = 0,
\end{eqnarray}
where the asterisk denotes the complex conjugate. Separating the perturbation into
its real and imaginary parts, according to $u_1 = X + iY$, and letting 
$X, Y \propto \exp(iKx - i\Omega t)$, we obtain from (2) the nonlinear dispersion relation
\begin{eqnarray}
  \Omega = -(Q_2 + 2Q_3)u_0^2K 
  \pm \left[ 
    Q_3^2u_0^4K^2 + (PK^2 - 4Q_1u_0^4)PK^2 
  \right]^{1/2}  ,
  \label{eq:dispersion}
\end{eqnarray}
where $K$ and $\Omega$ are the wavenumber and the frequency of low-frequency 
perturbations modulating the carrier signal. For $\Omega = i\Gamma  -(Q_2 + 2Q_3)u_0^2K$, 
we obtain the modulational instability growth rate from (3)
\begin{equation}\label{eq:mod}
  \Gamma = K\left[ (4PQ_1 - Q_3^2)u_0^4 - P^2K^2 \right]^{1/2},
\end{equation}
when $PQ_1 > 0$. We see that the effect of the derivative nonlinearity 
$Q_3$ is to decrease the instability region, while the higher order nonlinearity
coefficient $Q_1$ tends to increase the instability region. In Fig.\ 1, the typical 
structure of the modulational instability growth rate is depicted.


In order to analyze the effects due to partial coherence on the pulse
propagation in nonlinear electrical transmission lines, we next introduce the 
Wigner function, defined as the Fourier transform of the two-point 
correlation function \cite{Wigner}
\begin{equation}\label{eq:wigner}
  \rho(t,x,k) = \frac{1}{2\pi}\int\,d\xi\,e^{ik\xi}\langle 
    u^*(t,x + \xi/2)u(t,x - \xi/2) \rangle ,
\end{equation}
where the angular bracket denotes the ensemble average \cite{Klimontovich}. 
The Wigner function  defines a generalized phase space distribution 
function for quasi-particles, which satisfies the relation
\begin{equation}
  I(t,x) \equiv \langle|u(t,x)|^2\rangle = \int\,dk\,\rho(t,x,k) ,
\end{equation}
where $I$ is the pulse intensity.  Applying the time derivative on the 
definition (\ref{eq:wigner}) and using the MCGL equation (\ref{eq:nlse}),
we obtain \cite{Wigner,Moyal}
\begin{eqnarray}
  &&\!\!\!\!\!\!\!\!\!\!\!\!
  \partial_t\rho - 2Pk\partial_x\rho 
  - 2Q_1I^2\sin\left( \tfrac{1}{2}\stackrel{\leftarrow}{\partial}_x\stackrel{\rightarrow}{\partial}_k \right)\rho
  - Q_2I\left[
    \cos\left( \tfrac{1}{2}\stackrel{\leftarrow}{\partial}_x\stackrel{\rightarrow}{\partial}_k \right)\partial_x\rho
    - 2\sin\left( \tfrac{1}{2}\stackrel{\leftarrow}{\partial}_x\stackrel{\rightarrow}{\partial}_k \right)k\rho
  \right]
  \nonumber \\ &&
  - Q_3\left\{ 
    \partial_x\left[ 
      I\cos\left( \tfrac{1}{2}\stackrel{\leftarrow}{\partial}_x\stackrel{\rightarrow}{\partial}_k\right)\rho
    \right]
    -2kI\sin\left( \tfrac{1}{2}\stackrel{\leftarrow}{\partial}_x\stackrel{\rightarrow}{\partial}_k \right)\rho
  \right\} = 0 , 
  \label{eq:kinetic} 
\end{eqnarray}
where the $\sin$ and $\cos$ operators are defined in terms of their respective Taylor
expansion.  We note that the $\gamma$-term drops out, since it contains only
the phase information for $u$. Equation (\ref{eq:kinetic}) describes the propagation
of partially coherent pulses in nonlinear electrical transmission lines. 

We now analyze Eq.\ (\ref{eq:kinetic}) for small perturbations, i.e.\ 
we let $\rho(t,x,k) = \rho_0(k) + \rho_1\exp(iKx - i\Omega t)$ and 
$I(t,x) = I_0 + I_1\exp(iKx - i\Omega t)$, where $|\rho_1| \ll \rho_0$
and $|I_1| \ll I_0$. Linearizing Eq.\ (\ref{eq:kinetic}) we obtain the 
nonlinear dispersion relation
\begin{eqnarray}
  &&\!\!\!\!\!
  1 = \int dk\,\Bigg\{
    \frac{2Q_1I_0 + (Q_2 - Q_3)k - \tfrac{1}{2}(Q_2 + Q_3)K}{\Omega + 2PKk 
      +(Q_2 + Q_3)KI_0}\rho_{0-}
  \nonumber \\ && \qquad\qquad
    - \frac{2Q_1I_0 + (Q_2 - Q_3)k + \tfrac{1}{2}(Q_2 + Q_3)K}{\Omega + 2PKk 
      +(Q_2 + Q_3)KI_0}\rho_{0+}
    \Bigg\} ,
    \label{eq:gen-disp}
\end{eqnarray}
where $\rho_{0\pm} = \rho_0(k \pm K/2)$.


If the background wave function $u_0$ has a partially coherent  phase, the corresponding  
quasi-particle distribution is given by the Lorentzian \cite{Loudon}
\begin{equation}\label{eq:lorentz}
  \rho_0(k) = \frac{I_0}{\pi}\frac{\Delta}{k^2 + \Delta^2} ,
\end{equation}
where $\Delta$ is the width of the distribution, giving the degree of decoherence
of the pulse intensity. Using the distribution (\ref{eq:lorentz}) in the general
dispersion relation (\ref{eq:gen-disp}), we obtain 
\begin{eqnarray}
  &&\!\!\!\!\!\!\!\!\!\!\!\!
  \Omega = -(Q_2 + 2Q_3)I_0K + 2iP\Delta K 
  \nonumber \\ &&
  \pm \big[ 
    Q_3^2I_0^2K^2 + (PK^2 - 4Q_1I_0^2)PK^2 
    + 2iP(Q_2 - Q_3)I_0\Delta K^2 \big]^{1/2}  .
\end{eqnarray}
We will assume that $PQ_1 > 0$ in order to make a comparison to the coherent case.
With the normalization $\Omega = Q_1I_0^2\widetilde{\Omega}$, 
$K = (Q_1/P)^{1/2}I_0\widetilde{K}$, $\Delta = (Q_1/P)^{1/2}I_0\widetilde{\Delta}$,
$Q_2 = (PQ_1)^{1/2}\widetilde{Q}_2$, and $Q_3 = Q_2\widetilde{Q}_3$, we obtain
the dimensionless dispersion relation
\begin{eqnarray}
  \Omega = -Q_2(1 + 2Q_3)K + 2i\epsilon\Delta K 
  \pm \left[ 
    K^4 + (Q_2^2Q_3^2 -4)K^2  + 2i\epsilon Q_2(1 - Q_3)\Delta K^2
  \right]^{1/2}  ,
  \label{eq:norm-disp}
\end{eqnarray}
where we have dropped the tilde on all variables, and $\epsilon = \mathrm{sgn}(P)$.
In Fig.\ 1 we have plotted the normalized growth rate $\Gamma$ as a function of 
the normalized wavenumber $K$. We have assumed that $P < 0$, i.e.\ $\epsilon = -1$. 
When $Q_2 = Q_3/4=1/2$ and $\Delta = 0$, 
we obtain the full curve in Fig.\ 1, while $\Delta = 0.1$ gives the dashed curve. 
For $\Delta = 0.1$, but $Q_2= Q_3/4=-1/2$, we obtain the dotted curve in Fig.\ 1,
and $Q_3= Q_2/4=-1/2$ gives the dashed-dotted curve.  When $\epsilon =1$, a reduced 
distribution width $\Delta$ tend to increase the growth rate, which is unphysical. 

\begin{figure}
  \centering
  \includegraphics[width=0.9\columnwidth]{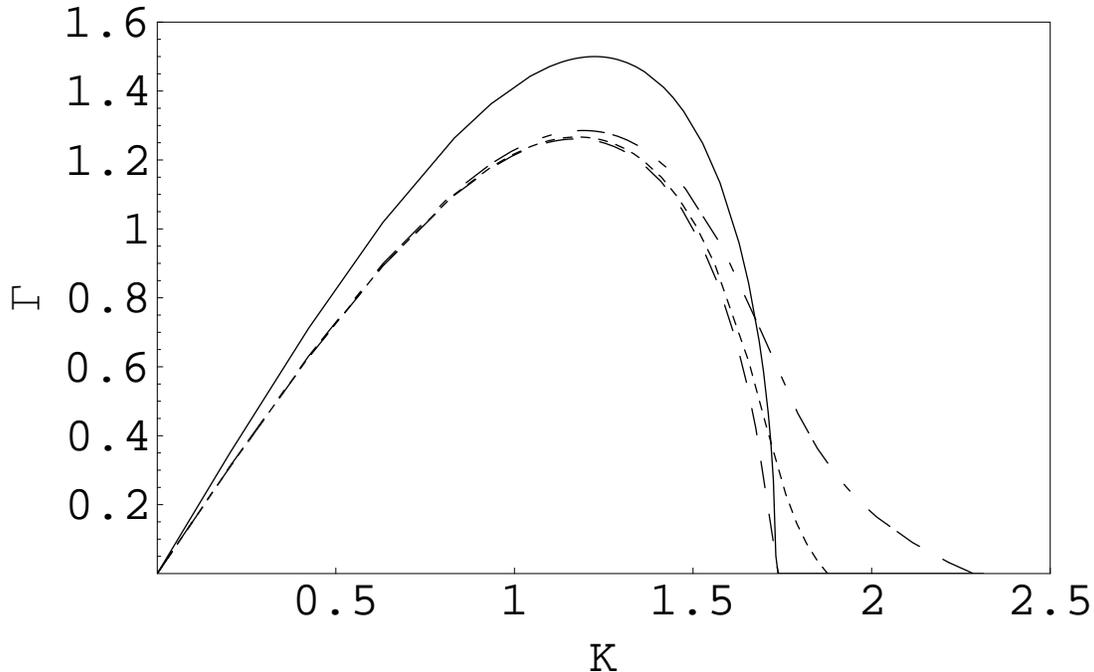}
  \caption{The typical structure of the modulational instability 
  growth rate $\Gamma$ as a function of the wavenumber $K$,
  as given by Eq.\ (\ref{eq:mod}). The full curve has full coherence 
  ($\Delta = 0$), while the dashed curve have a nonzero 
  decoherence as given by the width $\Delta$. The dotted curve 
  has negative values on the parameters $Q_2$ and $Q_3$ and 
  $Q_2= Q_3/4$, while $\Delta$ is still finite, and the dashed-dotted
  curve has $Q_3= Q_2/4 < 0$ with a finite $\Delta$.}
\end{figure}


To summarize, we have examined the modulational instability of 
partially coherent pulses in nonlinear electrical transmission lines.
For this purpose, we have derived a nonlinear dispersion relation
from the MCGL equation by using the Wigner-Moyal representation.
The nonlinear dispersion relation is analyzed for a Lorentzian 
equilibrium pulse distribution function. It is found that the
growth rate of the modulational instability is reduced due
to the consideration of the broadband signals. The present results 
should help to understand the nonlinear propagation of broadband 
pulses in electrical transmission lines.

\acknowledgments
M. M. thanks the members of the Institut f\"ur Theoretische
Physik IV and Centre for Plasma Science and Astrophysics 
at Ruhr-Universit\"at Bochum for their hospitality during his visit, 
when the present research was initiated. This research was partially 
supported by the Swedish Research Council.

\end{document}